 \documentclass[aps,prc,showpacs,floatfix,4paper]{revtex4}
\usepackage{amsmath}
\usepackage{amssymb}
\usepackage{epsfig}
\usepackage{graphicx}
\def\bfg #1{{\mbox{\boldmath $#1$}}}
\def\aphet{\bar p\,^3\rm{He}}
\def\aphef{\bar p\,^4\rm{He}}
\def\He{\rm{He}}

\begin{document}
\title{Antiproton scattering off $^3{\rm He}$ and $^4{\rm He}$ nuclei
 at low and intermediate energies}
 
\author {Yu.N. Uzikov$^{a}$, J. Haidenbauer$^{b,c}$, B.A. Prmantaeva$^{d}$}

\affiliation{
$^a$Laboratory of Nuclear Problems, Joint Institute for Nuclear
Research, 141980 Dubna, Russia\\
$^b$Institute for Advanced Simulation, Forschungszentrum J\"ulich,
D-52425 J\"ulich, Germany\\
$^c$Institut f\"ur Kernphysik and J\"ulich Center for Hadron Physics,
Forschungszentrum J\"ulich, D-52425 J\"ulich, Germany \\
$^d$L.N. Gumilyov Eurasian National University, 010008 Astana, Kazakhstan
}

\begin{abstract}
Antiproton scattering off $^3\He$ and $^4\He$ targets is considered at beam energies
below 300 MeV within the Glauber-Sitenko approach, utilizing the $\bar N N$ amplitudes of the
J\"ulich model as input. A good agreement with available data on 
differential $\aphef$ cross sections and on $\aphet$ and $\aphef$ reaction cross 
sections is obtained.
Predictions for polarized total $\aphet$ cross
sections are presented, calculated within the single-scattering approximation 
and including Coulomb-nuclear interference effects.
The kinetics of the polarization buildup is discussed.
%
\end{abstract}


\maketitle

\section{Introduction}
One of the projects suggested for the future FAIR facility in Darmstadt
comes from the PAX collaboration \cite{PAX}. Its aim 
is to measure the proton transversity in the interaction of polarized
antiprotons with protons.
In order to produce an intense beam of polarized antiprotons, the collaboration
intends to use antiproton elastic scattering off a polarized hydrogen
target ($^1$H) in a storage ring \cite{filtex}.
The basic idea is connected to the result of the FILTEX experiment \cite{FILT},
where a sizeable effect of polarization buildup was achieved in a
storage ring by scattering of unpolarized protons
off polarized hydrogen atoms at low beam energies of 23 MeV.
Recent theoretical analyses \cite{MS,NNNP,NNNP1,NNNP2} have shown that
the polarization buildup observed in Ref.~\cite{FILT} can be understood
quantitatively. According to those authors it is solely due to the spin
dependence of the hadronic (proton-proton) interaction which leads to the
so-called spin-filtering mechanism, i.e. to a different rate of removal
of beam protons from the ring for different polarization states of the
target proton.

In contrast to the $NN$ case, the spin dependence of the $\bar NN$
interaction is poorly known. Therefore, it is an open question whether
any sizeable polarization buildup can also be achieved in case of an
antiproton beam based on the spin-filtering mechanism. Indeed,
recently several theoretical studies were performed with the aim
to estimate the expected polarization effects for antiprotons,
employing different $\bar{p}p$ interactions
\cite{DmitrievMS,we09,DMS2}.
Besides of using polarized protons as target one could also use
light nuclei as possible source for the antiproton polarization
buildup. Corresponding investigations for antiproton scattering
on a polarized deuteron target were presented in Refs.~\cite{we09,we11,Sal11}. 
As was shown in Refs.~\cite{we09,we11} on
the basis of the Glauber-Sitenko theory {\cite{Glauber1,Glauber} with 
elementary {$\bar p N$} amplitudes taken from the J\"ulich $\bar N N$ models
\cite{Hippchen,Hippchen1,Mull,mine},
the $\bar p  d$ interaction could provide a comparable or even more effective 
way than the $\bar p p$ interaction to obtain polarized antiprotons. 
This conjecture can be checked at a planned experiment \cite{AD} at the
AD (Antiproton Decelerator) facility at CERN.
 
Yet another option could be the scattering of antiprotons off a polarized $^3\He$
target. Since the polarization of the $^3{\rm He}$ nucleus is carried mainly by the neutron, 
the $\bar p n$ amplitudes are expected to dominate the spin observables of this reaction.
In the present work we calculate spin-dependent cross sections
for the $\aphet$ interaction on the basis of an approach similar to that 
developed in Ref.~\cite{we09}. Experimental information on $\aphet$ scattering is
rather sparse \cite{Balestra,Bianconi}. Thus, in order to examine the validity of the 
employed Glauber-Sitenko approach \cite{Glauber,chizlesniak} at low and intermediate
energies we consider here also the $\aphef$ system where the PS179 collaboration 
has performed several measurements 
\cite{Bales84,Bales85,Bales87,Bales88,Bales89,Batusov,Bales93,Zenoni} at the
LEAR facility at CERN. In particular, we calculate differential 
cross sections for elastic scattering and compare them with data available at 
beam momenta of 200 MeV/c \cite{Bales93} and 600 MeV/c \cite{Batusov}.
As far as we know, this is the first time that those PS179 data are
analyzed within an approach that utilizes elementary $\bar N N$ amplitudes 
taken from a microscopic model of the $\bar N N$ interaction. 
Though a few investigations of $\aphet$ and $\aphef$ scattering have been
performed before \cite{lakshmatikov81,Bendiscioli} based on the Glauber-Sitenko
theory, none of them connects directly with amplitudes generated from
potential models that are fitted to $\bar N N$ data.  

The paper is structured as follows: 
In Sect. II some details of the
formalism are given. In particular, we define the amplitudes and their
relation to the cross sections and we provide the relation between
the amplitudes of the $\aphet$ system with those of the elementary
$\bar N N$ interaction within the single-scattering approximation.
Expressions required for the inclusion of the Coulomb interaction
are provided too. 
In Sect. III predictions for $\aphet$ and $\aphef$ are given, obtained 
within the Glauber-Sitenko approach. The results are compared with the
available data for those systems. 
In Sect. IV the polarization efficiency for $\aphet$ is studied. 
We introduce the pertinent quantities and then present and discuss the
numerical results. 
The paper closes with a short Summary. 

\section{Formalism}

\subsection{Forward elastic $\aphet$ scattering amplitude 
and total cross sections}

In order to calculate the total unpolarized and spin dependent $\aphet$
cross sections we use the optical theorem.
If $\hat F(0)$ is the operator of forward elastic scattering for $\aphet$ and
$\rho$ is the spin-density matrix of the $\aphet$ system
then the total cross section, $\sigma$,
is given by 
 \begin{eqnarray}
 \sigma=\frac{4\pi}{k_{\bar p\tau}}Im\frac{Tr\rho\hat F(0)}{Tr \rho} \ , 
  \label{opt}
 \end{eqnarray}
where $k_{\bar p\tau}$ is the modulus of the center-of-mass (c.m.) momentum 
in the $\aphet$ system. The spin-density matrix for the $\aphet$ system is
 \begin{eqnarray}
 \rho= \frac{1+{\bfg \sigma}_{\bar p} {\bf P}_{\bar p}}{2} \cdot
\frac{1+{\bfg \sigma}_{\tau} {\bf P}_{\tau}}{2},
\label{rho}
\end{eqnarray}
 where ${\bfg \sigma}_{\bar p} $ and ${\bfg \sigma}_{\tau}$ are Pauli 
matrices acting on the $\bar p$ and $^3\He$ spin states, respectively, 
and ${\bf P}_{\bar p}$ (${\bf P}_{\tau}$) is the polarization vector
of the antiproton ($^3\He$).
The operator $\hat F(0)$ for elastic scattering of two spin-$\frac{1}{2}$ particles
contains three terms \cite{bystricky}, 
\begin{eqnarray}
 \hat F(0)= F_0+F_1 {\bfg \sigma}_{\bar p}\cdot {\bfg \sigma}_{\tau}+
F_2({\bfg \sigma}_{\bar p}\cdot \hat {\bf k}) 
({\bfg \sigma}_{\tau\, }\cdot \hat {\bf k}),
\label{f0}
\end{eqnarray}
where $F_0,\, F_1, \, F_2$
are complex amplitudes and $\hat {\bf k}$ is the unit vector along the beam direction.
Inserting Eqs. (\ref{rho}) and (\ref{f0}) into Eq. (\ref{opt})
one  obtains
\begin{eqnarray}
 \sigma= \sigma_0+\sigma_1 {\bf P}_{\bar p}\cdot {\bf P}_{\tau}+
\sigma_2({\bf P}_{\bar p}\cdot \hat {\bf k}) 
({\bf P}_{\tau\, }\cdot \hat {\bf k}),
\label{sigmatot}
\end{eqnarray}
 where the total unpolarized cross section $\sigma_0$
 and the total spin-dependent
 cross sections $\sigma_1$ and $\sigma_2$ are introduced as
\begin{eqnarray}
\label{sigma0}
 \sigma_0= \frac{4\pi}{k_{\bar p\tau}} ImF_0,\\
\label{sigma1}
\sigma_1= \frac{4\pi}{k_{\bar p\tau}} ImF_1,\\
\label{sigma2}
\sigma_2= \frac{4\pi}{k_{\bar p\tau}} ImF_2.
\end{eqnarray}

\subsection{Single-scattering approximation}

 For the ground state  of the $^3\He$ nucleus we use the completely 
antisymmteric wave function $\Psi^A(1,2,3)$
 defined within the 
isospin formalism. Only the fully  symmetric spatial part,
 $\Psi_X^S$, and the antisymmetric spin-isospin part, $\xi^a$, are kept here \cite{ufn},
\begin{eqnarray}
\label{psiA}
 \Psi^A=\Psi_X^S\,\xi^a, \\ 
\xi^a=\frac{1}{\sqrt{2}}(\chi^\prime\zeta^{\prime \prime}-\chi^{\prime\prime}\zeta^\prime),
\label{chi}
\end{eqnarray}
where $\chi^\prime$, $\chi^{\prime\prime}$ are spin functions, and
$\zeta^\prime$, $\zeta^{\prime \prime}$ are those for the isospin.
For the $z$-projection of the $^3\He$ spin, $M_S=+\frac{1}{2}$, one has
the following spin wave functions, 
\begin{eqnarray}\label{chip1}
\chi^\prime&=&\frac{1}{\sqrt{2}}\alpha(1)[\alpha(2)\beta(3)-\beta(2)\alpha(3)], \\
\chi^{\prime \prime}&=&
\frac{1}{\sqrt{6}}\alpha(1)[\alpha(2)\beta(3)+\beta(2)\alpha(3)]-
\sqrt{\frac{2}{3}}\beta(1)\alpha(2)\alpha(3),
\label{chip2}
\end{eqnarray}
where $\chi^\prime$ is symmetric and  $\chi^{\prime\prime}$ is antisymmtric 
with respect to the permutation of the nucleons with the numbers 2 and 3.
In  Eqs. (\ref{chip1}) and (\ref{chip2}) the quantity 
$\alpha(i)$ ($\beta(i)$) corresponds to the eigenvalue of the $\sigma_z$-operator
$+1$ ($-1$) for the {\it i\,}th nucleon.
For the $^3\He$ spin projection $M_S=-\frac{1}{2}$ one should interchange 
$\alpha(1)$ and $\beta(1)$ in Eqs. (\ref{chip1}) and (\ref{chip2}), 
and replace $\alpha(2) \to  \beta(2)$, $\alpha(3) \to  \beta(3)$ in 
Eq.~(\ref{chip2}). 
The isospin wave functions $\zeta^\prime$ and $\zeta^{\prime \prime}$ are similar 
to those in Eqs. (\ref{chip1}) and (\ref{chip2}). 

In the single-scattering approximation the operator $\hat F$ of $\aphet$ scattering is taken 
within the isospin formalism as the following sum 
\begin{eqnarray}
\hat F=\frac{m_\tau}{m_N}\sqrt{\frac{s_{\bar p N}}{s_{\bar p \tau}}}[\hat f(1)+\hat f(2)+\hat f(3)],
\label{T123}
\end{eqnarray}
where the $\hat f(j)$'s (j=1,2,3) are operators in the $\bar p N$ spin-isospin space, 
\begin{eqnarray}
\hat f(j)=\frac{1}{2}(1+\tau_{z\,j}){\hat f}^{p}+\frac{1}{2}(1-\tau_{z\,j}){\hat f}^{n} \ .
\label{tj}
\end{eqnarray}
Here $m_N$ ($m_\tau$) is the mass of the nucleon ($^3\He$), $\sqrt{s_{\bar p N}}$
$(\sqrt{s_{\bar p \tau}})$ the invariant mass of the $\bar p N$  ($\bar p ~^3$He) 
system and ${\hat f}^{p}$ (${\hat t}^{n}$) is the operator related to 
$\bar p p$ ($\bar p n$) scattering with the same spin structure as given in Eq.~(\ref{f0}),
namely 
\begin{eqnarray}
 \hat f^N(0)= f_0^N+f_1^N {\bfg \sigma}_{\bar p}\cdot {\bfg \sigma}_{N}+
f_2^N({\bfg \sigma}_{\bar p}\cdot \hat {\bf k}) 
({\bfg \sigma}_{N }\cdot \hat {\bf k}),
\label{t0oper}
\end{eqnarray}
where $f_i$ (i=0,1,2) are complex amplitudes.
The matrix element of the operator $\hat F$ at zero scattering angle is
\begin{eqnarray}
<\sigma^\prime_{\bar p} M_S^\prime|F_{\aphet}|\sigma_{\bar p} M_S>=
3 \frac{m_\tau}{m_N}\sqrt{\frac{s_{\bar p N}}{s_{\bar p \tau}}}
<\Psi^s_x|\Psi^s_x>  \Bigl ( \frac{1}{6}<\chi^\prime|\hat f^p|\chi^\prime>+
\frac{1}{2}<\chi^{\prime\prime}|\hat f^p|\chi^{\prime\prime}>+
\frac{1}{3}<\chi^\prime|\hat f^n|\chi^\prime>\Bigr).
\label{m.e.}
\end{eqnarray}
Spin algebra gives from Eqs. (\ref{T123}), (\ref{tj}), (\ref{m.e.}) using
 (\ref{psiA}), (\ref{chi}), (\ref{chip1}) and  (\ref{chip2})
\begin{eqnarray}
\label{F012}
F_0= \frac{k_{\bar p \tau}}{k_{\bar p N}}(2f_0^p+f_0^n), 
\ \  F_1=- \frac{k_{\bar p \tau}}{k_{\bar p N}}f_1^n, \ \
 F_2= \frac{k_{\bar p \tau}}{k_{\bar p N}}(2f_1^n+f_2^n).
\end{eqnarray}
 Here $k_{\bar p N}$ is the c.m. momentum in the $\bar p N$ system 
which is related to the $\aphet$ momentum $k_{\bar p \tau}$ by 
\begin{eqnarray}
\frac{m_\tau}{m_N}\sqrt{\frac{s_{\bar p N}}{s_{\bar p \tau}}}=
\frac{k_{\bar p \tau}}{k_{\bar p N}},
\end{eqnarray}
which is valid for equal ($\bar p$) beam momenta in the $\bar p N$- and $\aphet$ systems. 
 One can see from Eq.~(\ref{F012}) that within the single-scattering 
 approximation the spin-dependent cross sections $\sigma_1$ and  
$\sigma_2$ are determined only by $\bar p$ scattering off
 the neutron. This result is in agreement with the fact that
 the matrix element of the operator of the
 $z$-projection of the $^3\He$ spin, $S_z$, written as 
\begin{equation}
S_z=\sum_{j=1}^{j=3} [s_z^p(j) \frac{1}{2}(1+\tau_{z}(j))+
s_z^n(j)\frac{1}{2}(1-\tau_{z}(j)]
\label{sz}
\end{equation}
 and sandwiched between the ground state wave function 
(\ref{psiA}) of $^3\He$, is completely determined  by the contribution of
the $z$-projection of the spin operator of the neutron, $s_z^n$, 
whereas the proton operator $s_z^p$ gives zero contribution: 
\hbox{$<\Psi^A_{M_S}|S_z|\Psi^A_{M_S}>=M_S$}, where $M_S=\pm \frac{1}{2}$.

 When substituting  Eqs. (\ref{F012}) into Eqs. (\ref{sigma0}), (\ref{sigma1}),
(\ref{sigma2}), one can find for the total $\aphet$ cross sections in 
single-scattering approximation (impulse approximation). 
 \begin{eqnarray}
\label{s0}
\sigma_0^{IA}&=&(2\sigma_0^{\bar p p}+\sigma_0^{\bar p n})\,{\widetilde w}, \\
\label{s1}
\sigma_1^{IA}&=& -\sigma_1^{\bar p n}\, {\widetilde w}, \\
\label {s2}
\sigma_2^{IA}&=&(2\sigma_1^{\bar p n}+\sigma_2^{\bar p n})\,{\widetilde w},
\label{s012}
\end{eqnarray}
where $ \widetilde w =<\Psi^s_x|\Psi^s_x> $. In the actual calculation we
set $<\Psi^s_x|\Psi^s_x> =1$.
The total $\bar p N$ cross sections $\sigma_i^{\bar p N}$ $(i=0,1,2)$ are determined
in Ref. \cite{bystricky} in the same way as in Eq. (\ref{sigmatot}) 
and their relations with the zero-angle elastic scattering
amplitudes $f_0,\, f_1,\, f_2$ can be found in Ref. \cite{bystricky}. 
Note that in Ref. \cite{MS} a different definition for the total polarized 
cross sections $\sigma_i$ (i=1,2) is used where then those quantities 
actually correspond directly to the transversal and longitudinal cross sections. 
Their cross sections ($\sigma_{i\,(MS)}$) are related to ours via $\sigma_1=\sigma_{1\,(MS)}$,
$\sigma_2=\sigma_{2\,(MS)}-\sigma_{1\,(MS)}$. Eqs. (\ref{s0}) and (\ref{s1}) are 
not changed when being rewritten in terms of $\sigma_{i\,(MS)}$, but 
Eq. (\ref{s2}) takes then the form $\sigma_{2\,(MS)}^{IA}=\sigma_{2\,(MS)}^{\bar p n}$. 
 
\subsection{Coulomb effects}

Coulomb effects are sizeable at low energies, i.e. for $T_{lab}\leq 25 $ MeV, 
as can be seen from the analysis of the FILTEX experiment \cite{filtex} 
in which protons were scattered off polarized hydrogen at 23 MeV.
For $\aphet$ scattering Coulomb effects could be even more important due
to the twice-as-large electric charge of $^3\He$.

The Coulomb amplitude of elastic $\aphet$ scattering is \cite{LL}
\begin{eqnarray}
f_c(\theta)=-\Bigl (\frac {\eta} {2 k_{\bar p \tau}\sin^2{(\theta/2)}}\Bigr )
\exp{\{i\eta \ln{\sin^{-2}(\theta/2)}+2i\tilde{\sigma}_0\} }.
\label{fc}
\end{eqnarray}
Here $\eta =Z_1Z_2\alpha \mu_{\bar p \tau}/k_{\bar p \tau}$ with $Z_1Z_2=-2$,
 $\alpha$ is the fine structure constant and $\mu_{\bar p \tau}$ is the reduced mass
of the $\aphet$ system. The Coulomb phase is given by
$\tilde {\sigma}_0=\arg\Gamma(1+i\eta)$, where $\Gamma(z)$ is the gamma function.

The total unpolarized Coulomb cross section $\sigma_0^C$ is estimated here following 
Ref. \cite{MS}, where proton-proton scattering in storage rings was analyzed. It leads to
the following result:
 \begin{eqnarray}
 \sigma_0^C=\pi \Bigl (
\frac{4\alpha \mu_{\bar p \tau}} {k_{\bar p \tau}^2\theta_{acc}}\Bigr )^2.
\label {couls0}
\end{eqnarray}
Here $\theta_{acc}<< 1$ is the beam acceptance angle, which is defined so that
for scattering at smaller angles $\theta \le \theta_{acc}$ the antiprotons remain 
in the beam. 
 The polarized total Coulomb cross sections $\sigma_1^C$ and  $\sigma_2^C$
 are zero for $\aphet$ scattering, since the nonrelativistic 
 Coulomb elastic scattering amplitude does not depend on the spins of $\bar p$ and $^3\He$
 and, in contrast to $pp$ scattering, does not contain antisymmetrization terms.
 The remaining part of the Coulomb effects is related to Coulomb-nuclear interference.
 The spin structure of the $\aphet$ scattering amplitude is similar to that
 for $pp$ scattering. Therefore, the cross sections due to the interference terms, 
 $\sigma_0^{int}$, $\sigma_1^{int}$,
 and $\sigma_2^{int}$, are calculated here on the basis of the formalism developed in
 Refs. \cite{MS,we09}. The final result for $\aphet$ can be obtained from the one
for ${\bar p}p$ scattering given in Eq. (27) of Ref. \cite{we09} 
 via the following substitutions: 
 $\alpha \to 2\alpha$,
 $m_p/2\to \mu_{\bar p \tau}$, $\chi_0\to \tilde {\sigma_0}$. Furthermore,  
the zero-angle helicity
amplitudes $M^p_i(0)$ (i=1,2,3) of the hadronic $\bar p p$ scattering
have to be replaced by the corresponding helicity amplitudes of 
zero-angle $\aphet$ scattering, $M_i^{\tau}(0)$. 
When using the single-scattering approximation given by Eqs. (\ref{F012}),
one finds the following expressions for the contribution of the Coulomb-nuclear 
interference terms to the total cross sections, 
\begin{eqnarray}
\nonumber
\sigma_0^{int} =&&-\frac{2\pi}{k_{\bar pN}}
\Bigl \{
 \cos{2\tilde {\sigma}_0}\bigl [-\sin{\Psi} Re  {\tilde M}_0
 +(1-\cos{\Psi})Im {\tilde M}_0\bigr ] - \\
\nonumber 
&&-\sin{2\chi_0}\bigl [\sin{\Psi} Im {\tilde M}_0
 +(1-\cos{\Psi})Re {\tilde M}_0\bigr ] \Bigr \},\\
\nonumber
\sigma_1^{int} =&& -\frac{2\pi}{k_{\bar pN}}
\Bigl \{
\cos{2\tilde{\sigma}_0} \bigl [ \sin{\Psi} Re M_2^n(0)
-(1-\cos{\Psi})Im M_2^n(0) \bigr ] + \\ 
\nonumber 
&&+\sin{2\chi_0}\bigl [ \sin{\Psi} Im M_2^n(0)
 +(1-\cos{\Psi})ReM_2^n(0)\bigr ]
\Bigr \}, \\
\label{CNpp2}
\sigma_2^{int}=&&-\frac{2\pi}{k_{\bar pN}}
\Bigl \{
\cos{2\chi_0}\bigl [-\sin{\Psi} Re {\tilde M}_2 
 +(1-\cos{\Psi})Im {\tilde M}_2 \bigr ] + \nonumber \\ 
&&+\sin{2\chi_0}\bigl [\sin{\Psi} Im {\tilde M}_2
 -(1-\cos{\Psi})Re {\tilde M}_2\bigr ]
\Bigr 
\}, 
\end{eqnarray}
where the following notations are used,
\begin{eqnarray}
{\tilde M}_0&=&2M_1(0)^p+2M_3(0)^p +M_1^n(0)+M_3^n(0), \nonumber \\
{\tilde M}_2&=&M_2(0)^n+M_3(0)^n -M_1^n(0),  \nonumber \\
\Psi&=&2\eta\ln{\sin{\theta_{acc}/2}}.
\end{eqnarray}

\begin{figure}[t]
 \includegraphics[width=0.40\textwidth,angle=0]{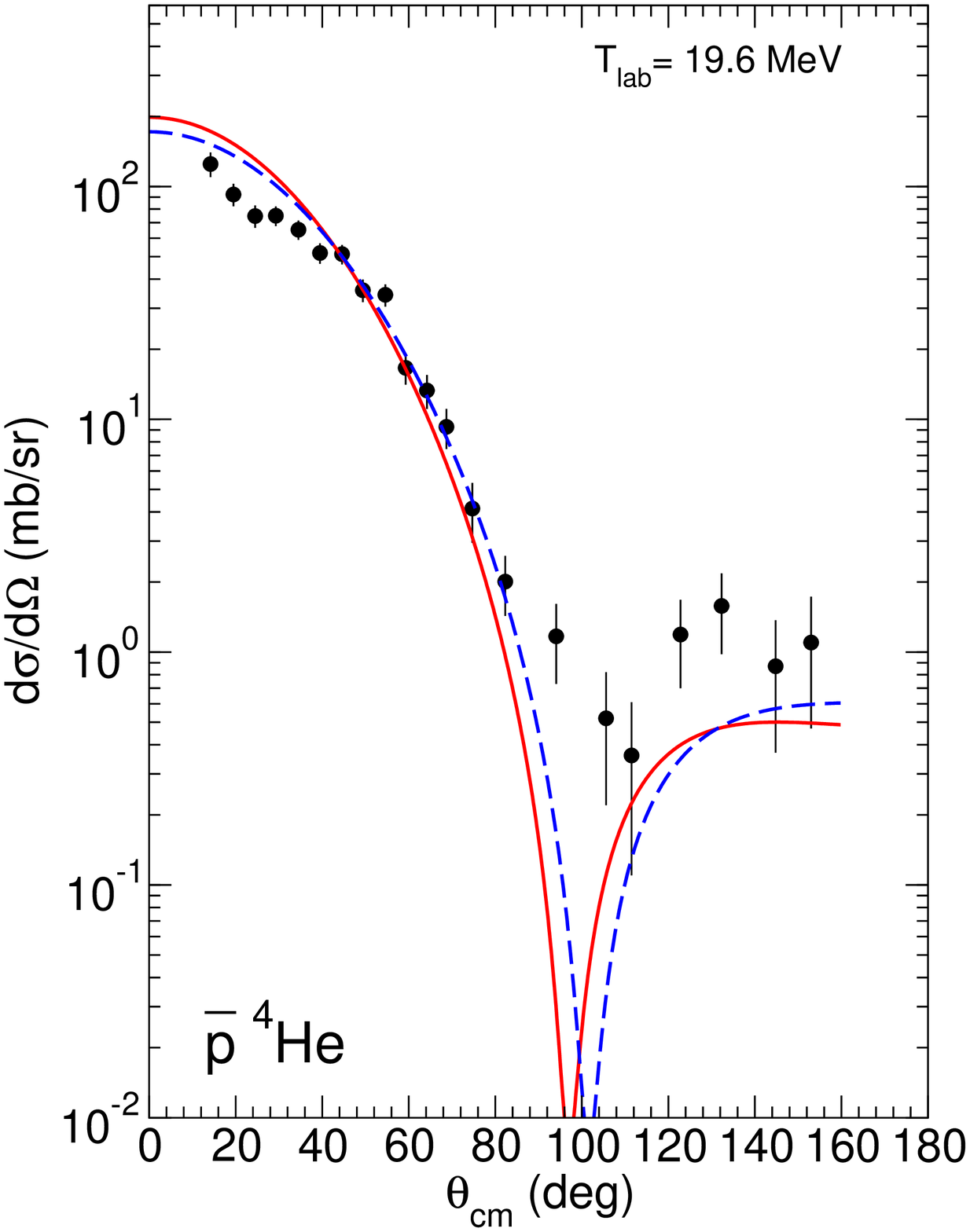}\includegraphics[width=0.40\textwidth,angle=0]{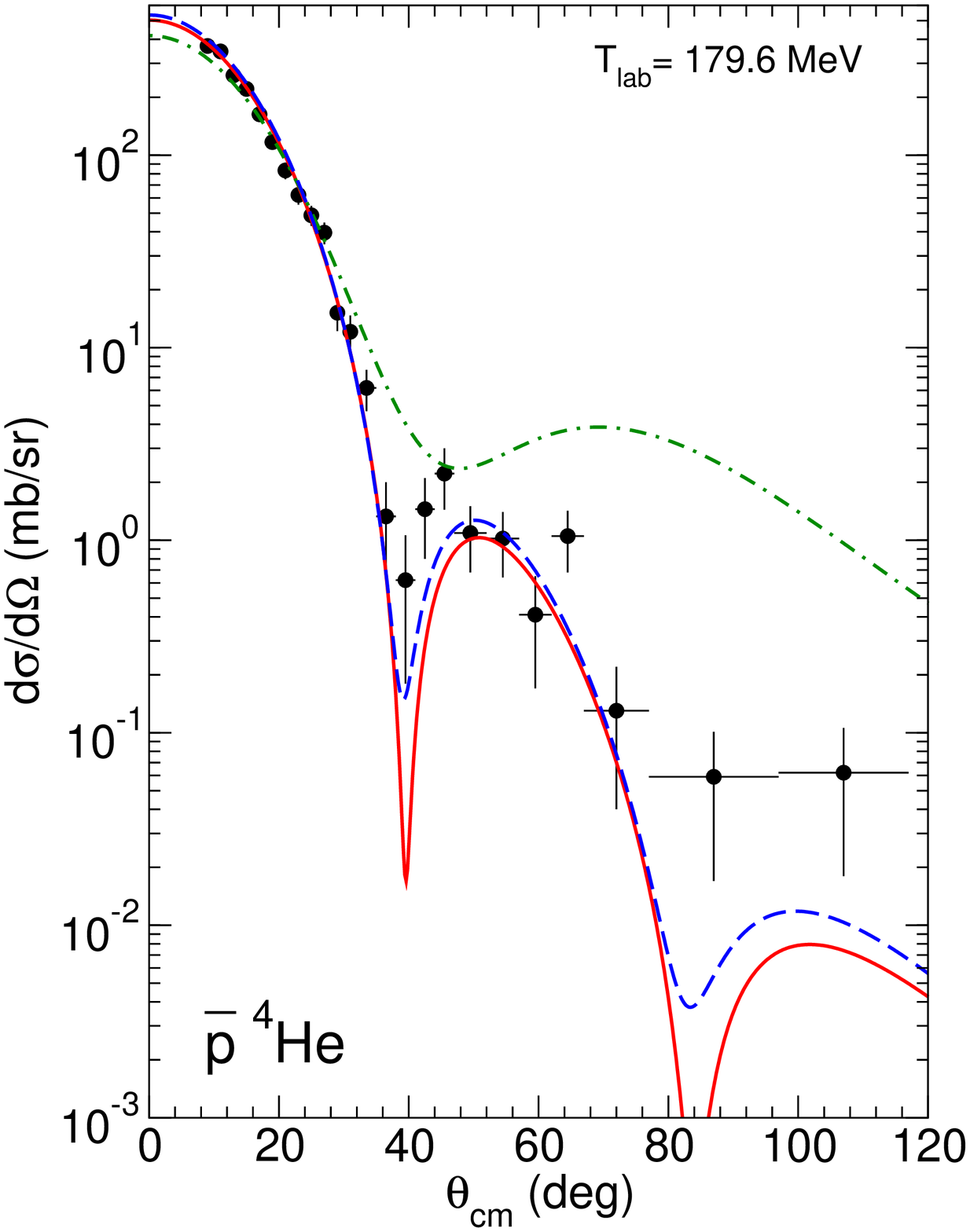}
\caption{Differential cross section for $\aphef$ versus the c.m. scattering angle
at $T_{lab}=$ 19.6 MeV and 179.6 MeV. The solid and dashed lines are results 
for the $\bar NN$ models D and A, respectively, obtained on
the basis of the approach \cite{chizlesniak}. 
The dash-dotted line is the result obtained within the approximation \cite{hasell} 
for the J\"ulich model D. 
Data are taken from Refs.~\cite{Bales93} (19.6 MeV) and \cite{Batusov} (179.6 MeV).
}
\label{difpalpha}
\end{figure}

\section{Results for $\aphet$ and $\aphef$ based on the Glauber-Sitenko approach}

In the present investigation we use two $\bar N N$ models
developed by the J\"ulich group.
Specifically, we use the models A(BOX) introduced in Ref.~\cite{Hippchen}
and D described in Ref.~\cite{Mull}.
Starting point for both models is the full Bonn $NN$ potential~\cite{MHE};
it includes not only traditional one-boson-exchange diagrams but also
explicit $2\pi$- and $\pi\rho$-exchange processes as well as virtual
$\Delta$-excitations. The G-parity transform of this meson-exchange
$NN$ model provides the elastic part of the considered $\bar N N$ interaction
models.
In case of model A(BOX) \cite{Hippchen} (in the following
referred to as model A)
a phenomenological \hbox{spin-}, isospin- and energy-independent
complex potential of Gaussian form is added to account for the
$\bar N N$ annihilation. It contains only three free parameters (the range
and the strength of the real and imaginary parts of the annihilation
potential), fixed in a fit to the available total and integrated
$NN$ cross sections.
In case of model D \cite{Mull}, the most complete $\bar N N$ model of the
J\"ulich group, the $\bar N N$ annihilation into 2-meson decay
channels is described microscopically, including all possible
combinations of $\pi$, $\rho$, $\omega$, $a_0$, $f_0$, $a_1$, $f_1$,
$a_2$, $f_2$, $K$, $K^+$ -- see Ref. \cite{Mull} for details --
and only the decay into multi-meson channels is simulated by
a phenomenological optical potential.
Results for the total and integrated elastic ($\bar p p$) and
charge-exchange ($\bar p p \to \bar nn$) cross sections and also
for angular dependent observables for both models can be found in
Refs.~\cite{Hippchen,Mull,mine}. Evidently, with model A as well as 
with D a very good overall reproduction of the low- and intermediate
energy $\bar N N$ data was achieved.

The unpolarized cross sections for $\aphet$ and $\aphef$ are calculated 
using the multiple scattering theory of Glauber-Sitenko \cite{Glauber,chizlesniak}. 
It is known that for proton scattering on nuclei this theory is only valid 
at fairly high energies, say for energies from $\sim 1$ GeV upwards. 
This is different in case of the antiproton-nucleus interaction. 
Strong annihilation effects in the elementary ${\bar p}N$ interaction 
lead to a peaking of the ${\bar p}N$ elastic scattering amplitude in 
forward direction already at very low energies and, therefore, render it 
suitable for application of the eikonal approximation, which is the basis of the 
Glauber-Sitenko theory. As a consequence, for 
antiproton reactions this theory can be applied at much lower energies, 
namely $ \sim 50$ MeV or even less \cite{lakkolybas}. For example, 
for $\bar p d$ scattering we found that the Glauber-Sitenko theory
even seems to work at $T_{lab}\sim 25$ MeV \cite{we09}. 
However, since the radii of $^3\He$ and $^4\He$ are smaller than that of 
the deuteron, it is possible that for $\aphet$- and especially for
$\aphef$ scattering the onset of
applicability of the Glauber-Sitenko theory could occur at
somewhat higher energies. Thus, in order to explore the reliability 
of this theory it would be desirable to confront our results with 
experimental information. 
Unfortunately, for $\aphet$ the only published experimental result 
in the considered energy region is a $\aphet$ reaction cross section 
at the beam energy of 19.6 MeV \cite{Balestra}. 
There is one more data point, namely the $\aphet$ annihilation cross 
section close to threshold \cite{Bales89}, but this is certainly outside 
of the region where the Glauber-Sitenko theory can be used. 

Indeed, the experimental situation for $\aphef$ is much better. In this 
case the PS179 collaboration has published results for integrated 
\cite{Bales84,Bales85,Bales87,Bales88,Bales89} as well
as differential cross sections \cite{Batusov,Bales93}. Thus, 
as a test we performed also calculations for this system within
the Glauber-Sitenko approach.
In those calculations we employ a Gaussian representation of 
the $\bar p N$ scattering amplitude in the form
\begin{eqnarray}
\label{fpn}
  f_{\bar pN}(q)=\frac{k_{\bar pN}\sigma^{\bar p N}_{tot}(i+\alpha_{\bar p N})}
{4\pi} \exp{(-\beta^2_{\bar p N}q^2/2)},
\end{eqnarray}
where $q$ is the transferred 3-momentum. 
The parameters $\sigma^{\bar p N}_{tot}$, $\alpha_{\bar p N}$ and 
$\beta^2_{\bar p N}$
are fixed from the spin-averaged amplitudes $f_{\bar pN}$ of
the models A and D and given in Ref. \cite{we09}. 
We utilize the formalism of Ref.~\cite{chizlesniak}, where a Gaussian 
nuclear density is used and corrections from the c.m. motion are included.
Furthermore, we take into account explicitly that the $\bar p p$ and
$\bar p n$ scattering amplitudes are different. 
We adopt the nuclear radius $r=1.37$ fm 
for $^4\rm{He}$ \cite{chizlesniak} and 1.5 fm for $^3\rm{He}$ \cite{hasell}.
The differential cross section we
obtained for $\aphef$ scattering at 179.6 MeV is in rather good 
agreement with the data of Ref. \cite{Batusov} (see Fig.~\ref{difpalpha}). 
We want to emphasize that no free parameters are involved in our calculation. 
For comparison we examined also the formalism
of Ref. \cite{hasell} where the $\bar pN$ scattering amplitudes are 
 evaluated exactly for the single scattering mechanism, but 
taken out of the loop integrals for $pN$ ($\bar p$N) re-scattering of higher order.
 This approximation works rather well for proton-$^3\He$ 
scattering at a few hundred MeV \cite{hasell}, but in case of $\aphef$ 
scattering at 179.6 MeV its applicability seems to be limited to much smaller
scattering angles ($\theta_{cm}< 30^o$) as compared to the approach of
Ref. \cite{chizlesniak}, as is demonstrated in Fig.~\ref{difpalpha} (cf. 
the dash-dotted curve). 

Results at 19.6 MeV are also shown in Fig.~\ref{difpalpha} and compared
with experimental information from \cite{Bales93}. Obviously even at this fairly 
low energy, corresponding to a beam momentum of $p_{lab}=192.8$ MeV/c, the data 
are remarkably well reproduced. There is, however, an overestimation of
the differential cross section at very forward angles. 
We included the Coulomb amplitude given by Eq.(\ref{fc}) in addition to the
 hadronic   Glauber-Sitenko $\bar p ~^3He$ amplitude and found that at 19.6 MeV
 and scattering angles $\theta_{cm}$ less than $\approx 2^\circ$ the Coulomb
 contribution
 is important, but negligible at larger angles $\theta_{cm}> 5^\circ$ and therefore
 does not allow one to explain the observed deviation in forward direction at 
$20^\circ-40^\circ$.

\begin{figure}
\includegraphics[width=0.40\textwidth]{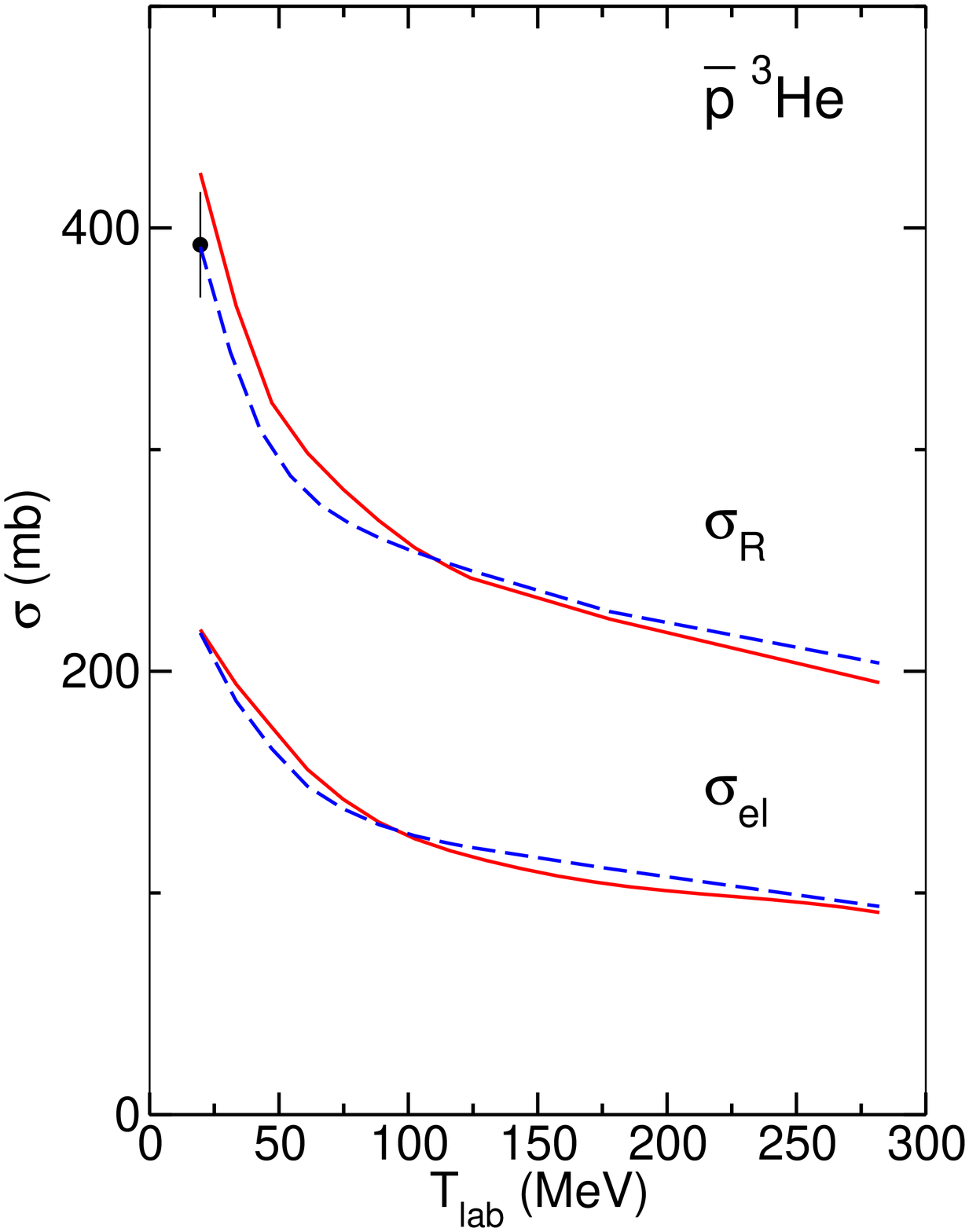}\includegraphics[width=0.40\textwidth]{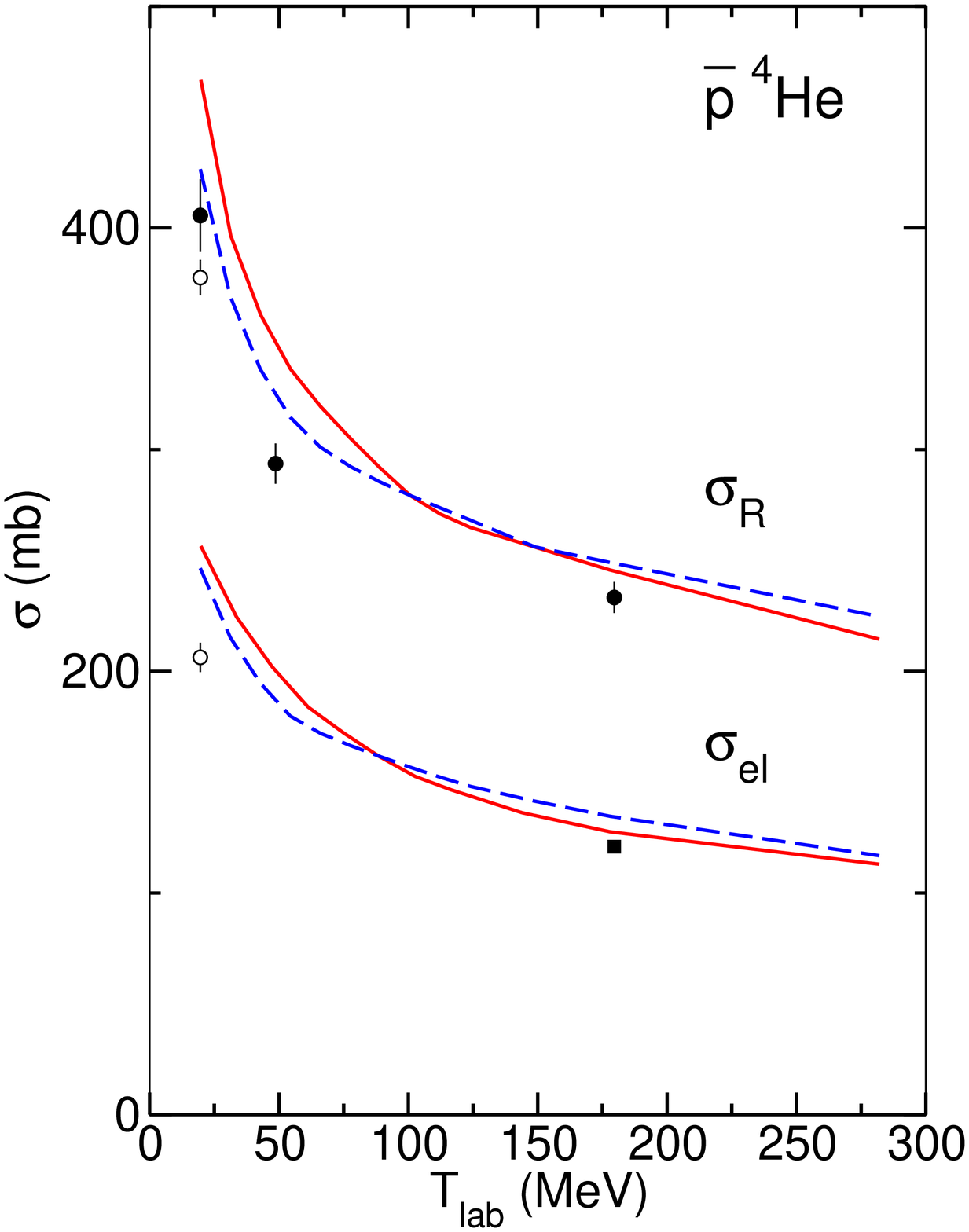}
\caption{Integrated elastic ($\sigma_{el}$, lower curves) and 
total reaction ($\sigma_{R}$, upper curves) cross sections for 
$\aphet$ and $\aphef$ versus the beam kinetic energy $T_{lab}$. 
The solid and dashed lines are results 
for the $\bar NN$ models D and A, respectively,  
obtained on the basis of the Glauber-Sitenko approach \cite{chizlesniak}.
Data for $\aphef$ are taken from Refs.~\cite {Bales85} 
(filled circles), \cite{Batusov} (squares), and \cite {Bales93} (open circles).
The data point for $\aphet$ is taken from Ref.~\cite {Balestra}. 
}
\label{Total}
\end{figure}

The total cross section can be evaluated by using the optical theorem. 
At $T_{lab}=19.6$ MeV where the $\aphet$ reaction cross section was
measured by the PS179 collaboration \cite{Balestra} we obtain 
$\sigma_0= 609$ mb for model A and 644 mb for model D.
Evaluating the differential cross section for elastic $\aphet$ scattering
allows us to compute also the integrated elastic cross section $\sigma_{el}$. 
Here we find $\sigma_{el}=$ 217 mb (A) and 219 mb (D). 
The reaction cross section is then given by
$\sigma_{R} = \sigma_0 - \sigma_{el}$ 
(we adopt here the notation of \cite{Bales87}).
Thus, we get 392 mb for model A and 425 mb for model D. 
The experimental result is 392$\pm$23.8 mb \cite{Balestra}.
It is quite remarkable that the Glauber-Sitenko 
theory combined with the J\"ulich models for the $\bar pN$ interaction
agrees so well with the measurement at this low energy. 

For $\aphef$ scattering experimental results for the reaction
cross section \cite{Bales87} as well as for the integrated elastic cross 
section \cite{Batusov,Bales93} have been published. Those data
points are displayed in Fig.~\ref{Total}, together with the
predictions of our calculations. One can see from the figure that
the model results are well in line with the energy dependence exhibited
by the data. But, in general, they overestimate the measured cross
sections by 5 to 10 \% (model A) and 10 to 20 \% (model D). 
In case of $\aphet$, also shown in 
Fig.~\ref{Total}, the predictions for both considered $\bar NN$ models 
agree with the experiment within the error bars, as was already
pointed out above. 

For completeness, predictions for the differential cross section for 
$\aphet$ scattering at two energies are displayed in Fig.~\ref{difpt}.
The results are qualitatively rather similar to those for the 
$\aphef$ system. 
 
\begin{figure}[t]
 \includegraphics[width=0.40\textwidth,angle=0]{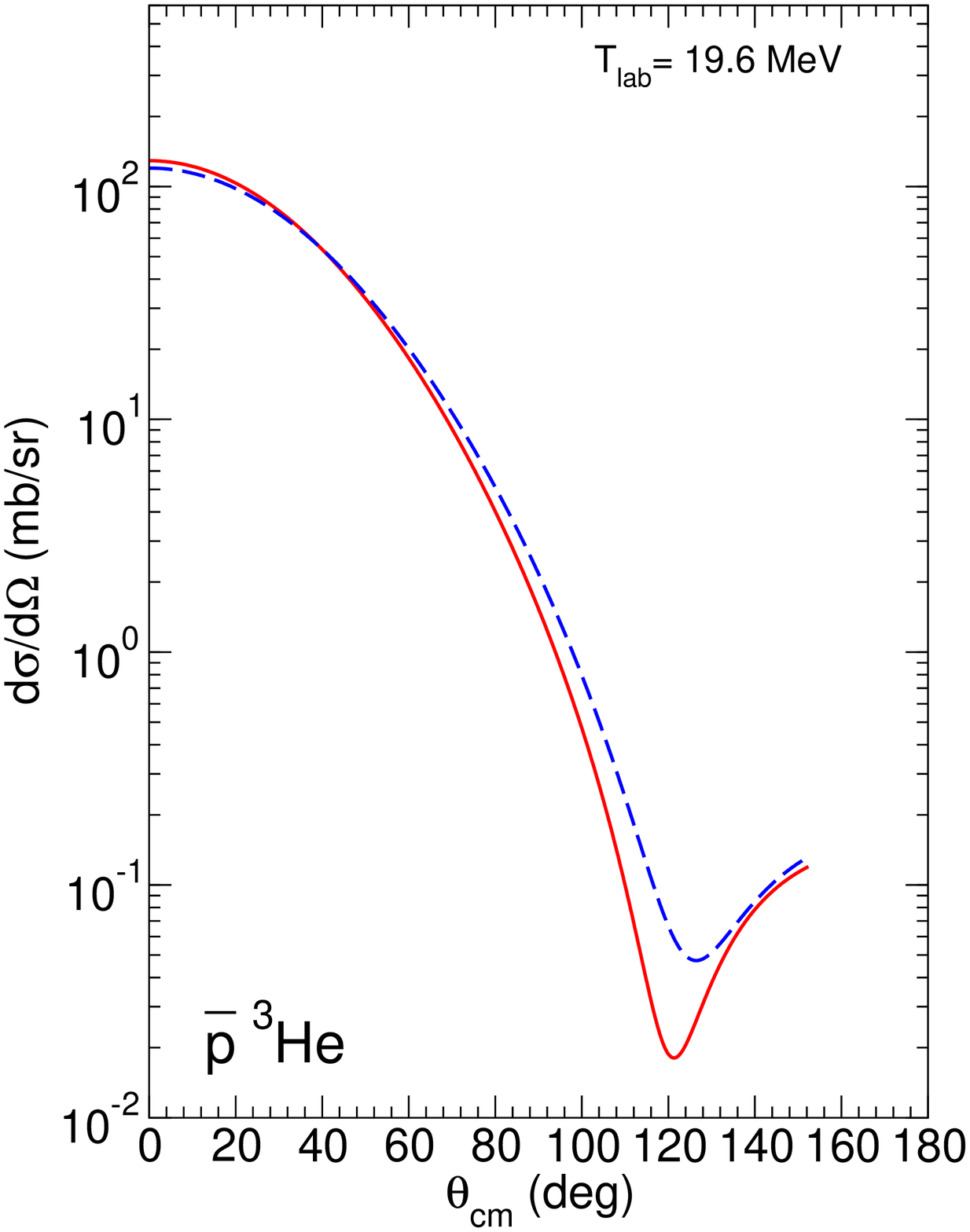}\includegraphics[width=0.40\textwidth,angle=0]{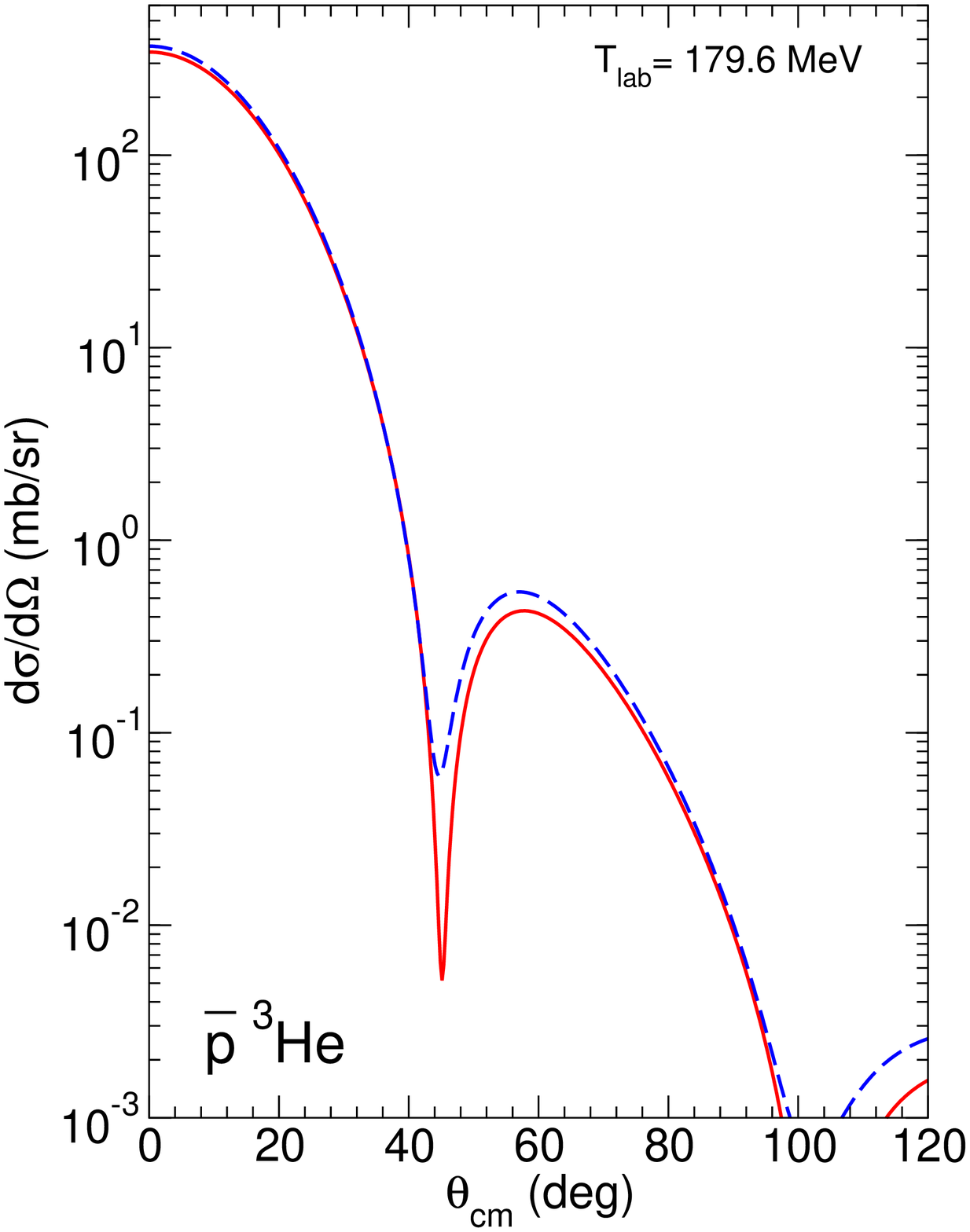}
\caption{Differential cross section for $\aphet$ versus the c.m. scattering angle
at $T_{lab}=$ 19.6 MeV and 179.6 MeV. The solid and dashed lines are results 
for the $\bar NN$ models D and A, respectively, obtained on
the basis of the approach \cite{chizlesniak}. 
}
\label{difpt}
\end{figure}

Finally, let us discuss here the so-called shadowing effects,
i.e. the corrections that arise in the multiple 
scattering approach of Glauber-Sitenko as employed in our calculation of 
the $\aphet$ and $\aphef$ scattering observables presented above.  
To determine the magnitude of the $\bar p N$ multiple 
scattering contributions quantitatively let us consider
the ratio of the total $\aphet$ cross section obtained within the 
single-scattering approximation to the one accounting for all
allowed 
orders of re-scattering, $R=\sigma_0^{IA}/\sigma_0$. 
We found that this ratio is roughly $1.45$ at low energies $\sim 25$ MeV 
and smoothly decreases to $R=1.33$ when the beam energy is increased to 179.6 MeV.
For $\ap d$ scattering this ratio was found to be smaller, namely
$\sim 1.1-1.15$ \cite{we09}.
The reason for this difference is the more compact structure of the $^3\He$ 
as compared to the loosely bound deuteron, which leads to an increase of the 
shadowing effects. Indeed this can be easily verified by simply increasing the 
radius of the Gaussian density $r$ to 4 fm in our calculation. 
Then the ratio $R$ smoothly reduces to 1.15 at 19.6 MeV and 1.09
at 179.6 MeV. 

\section{Polarized cross sections for $\aphet$ }

\begin{figure}
\includegraphics[width=0.30\textwidth]{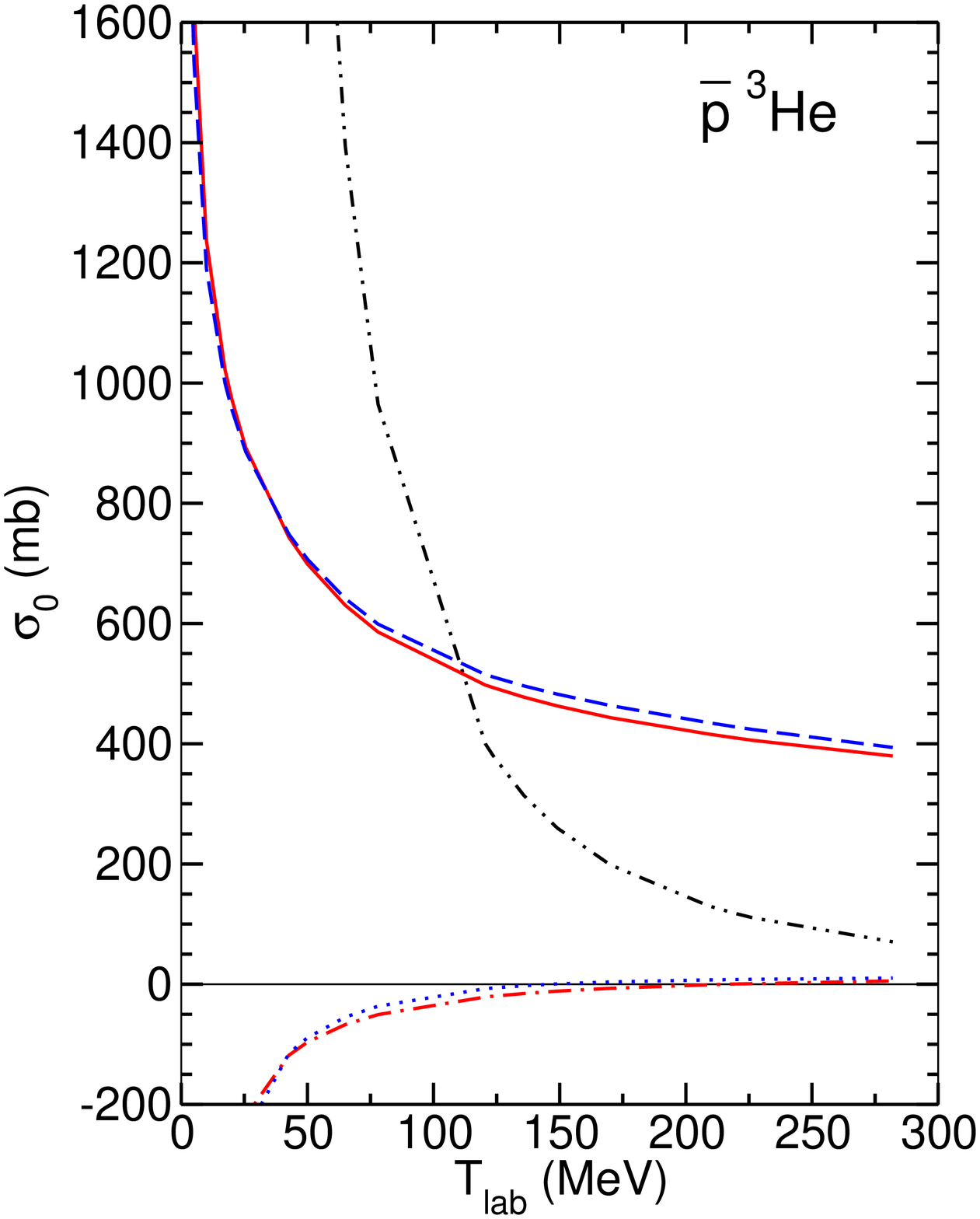}
\includegraphics[width=0.30\textwidth]{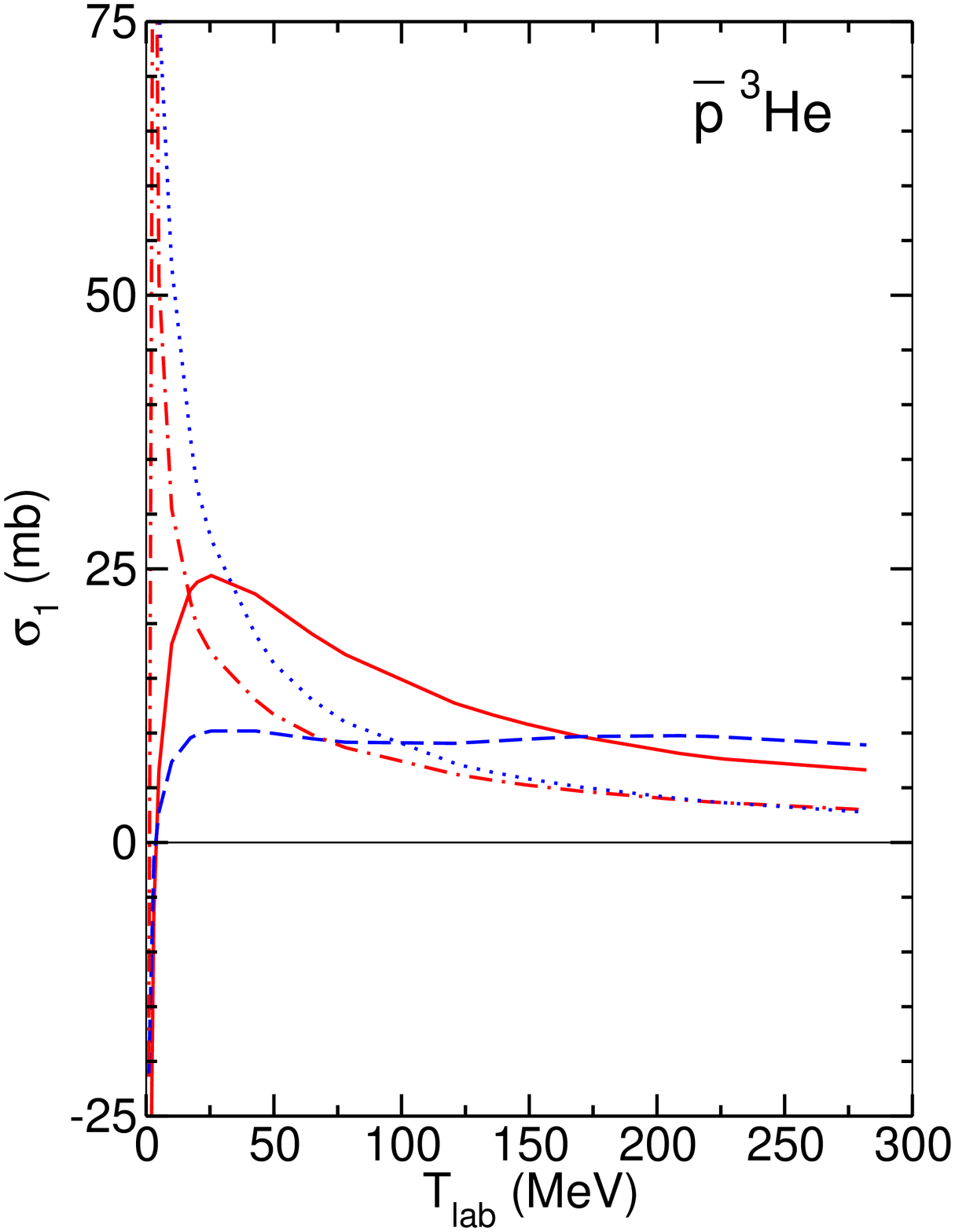}\includegraphics[width=0.30\textwidth]{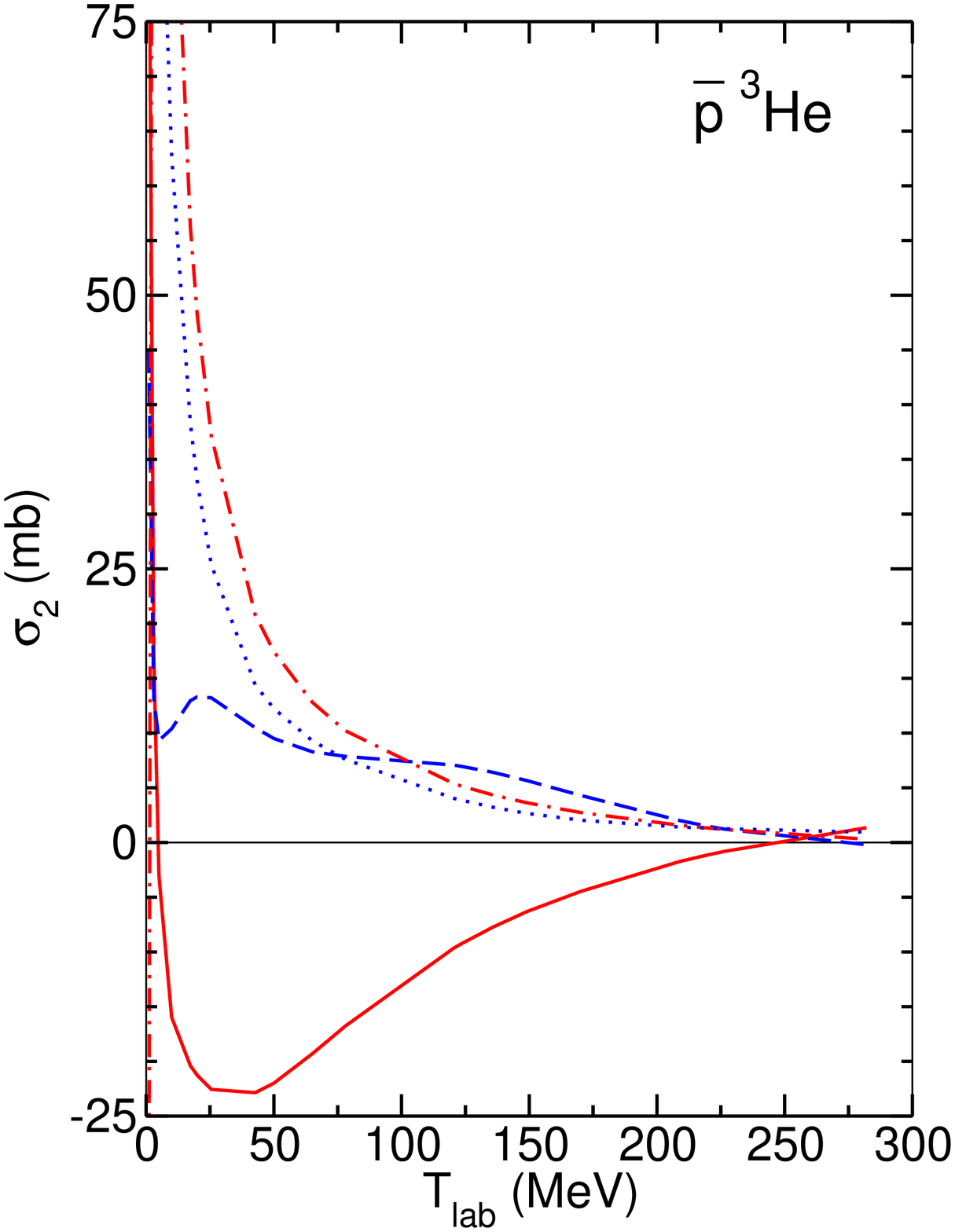}
\caption{
Total cross sections $\sigma_0$, $\sigma_1$
and $\sigma_2$ versus the antiproton laboratory energy $T_{lab}$
for $\aphet$ scattering. 
Results based on the purely hadronic amplitude, $\sigma^h_i$, 
(model D: solid line, model A: dashed line)
and for the Coulomb-nuclear interference term, 
$\sigma^{int}_i$, (D: dash-dotted line, A: dotted line), are presented.
In case of $\sigma_0$ the Coulomb cross section (cf. Eq.~(\ref{couls0})) is 
shown too (dash-double-dotted line).
The employed acceptance angle is $\theta_{acc}=10$ mrad. 
}
\label{Totals}
\end{figure}

\begin{figure}
 \includegraphics[width=0.30\textwidth,angle=0]{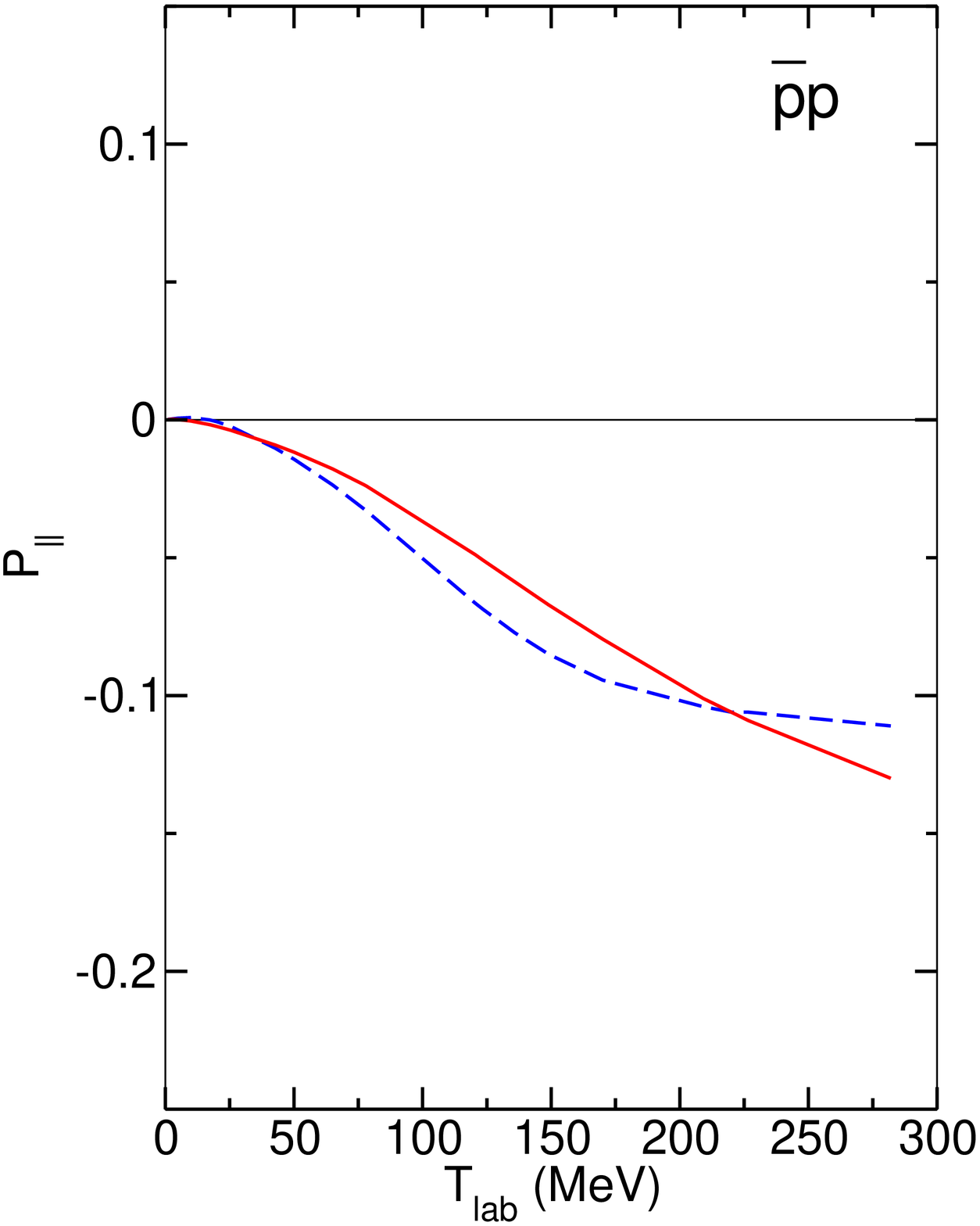}
\includegraphics[width=0.30\textwidth,angle=0]{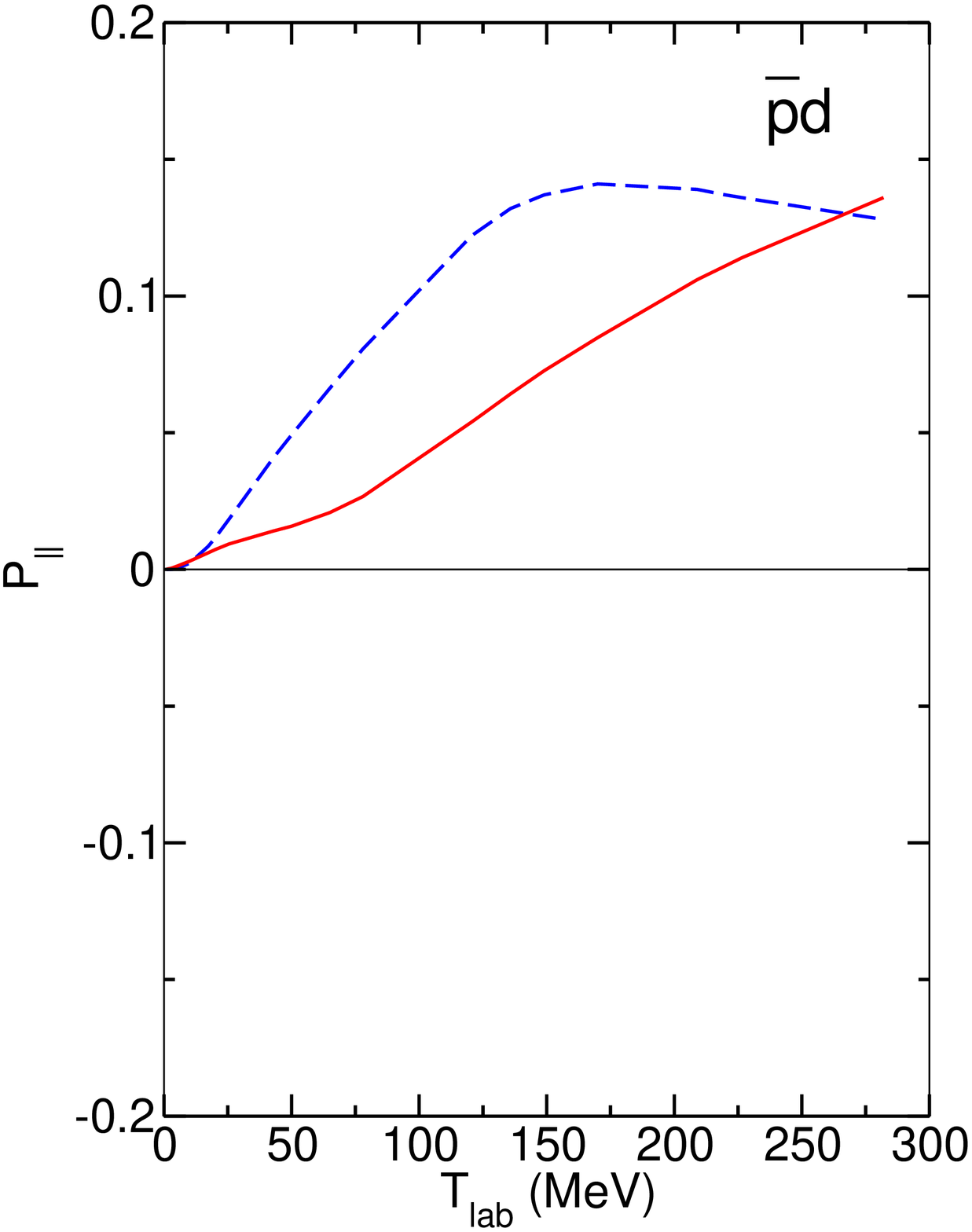}
\includegraphics[width=0.30\textwidth,angle=0]{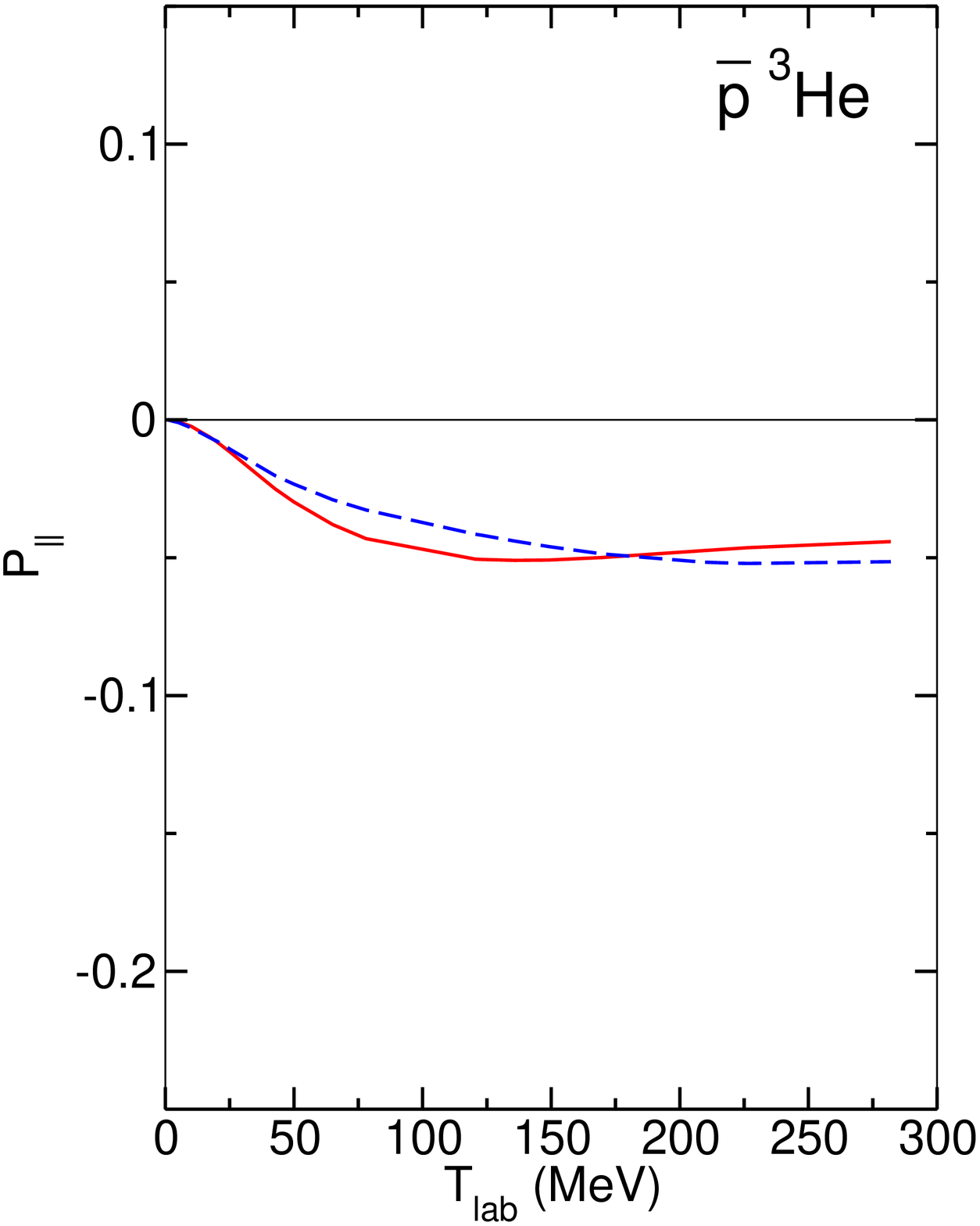}
\caption{
Dependence of the
longitudinal polarization $P_{||}$ (i.e. $P_{\bar p}(t_0)$ for
${\bfg \zeta}\cdot {\bf \hat k}=1$)
on the beam energy
for the target polarization  $P_T=1$ in the different reactions
$\bar pp$, $\bar pd$, and $\aphet$.
The results are for the models A (dashed line) and D (solid line).
The employed acceptance angle is $\theta_{acc}=10$ mrad. 
}
\label{ppara}
\end{figure}

\begin{figure}
 \includegraphics[width=0.30\textwidth,angle=0]{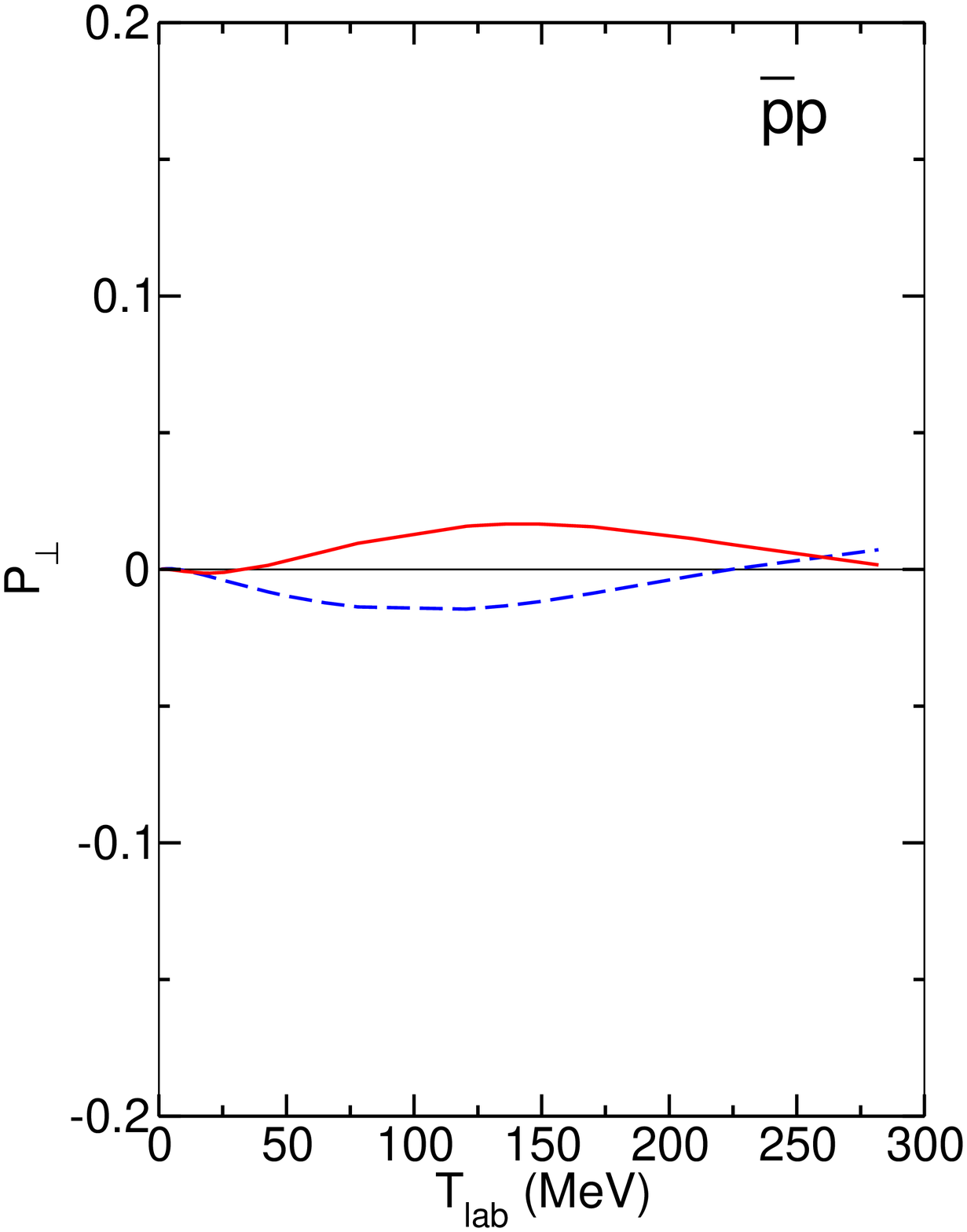}
\includegraphics[width=0.30\textwidth,angle=0]{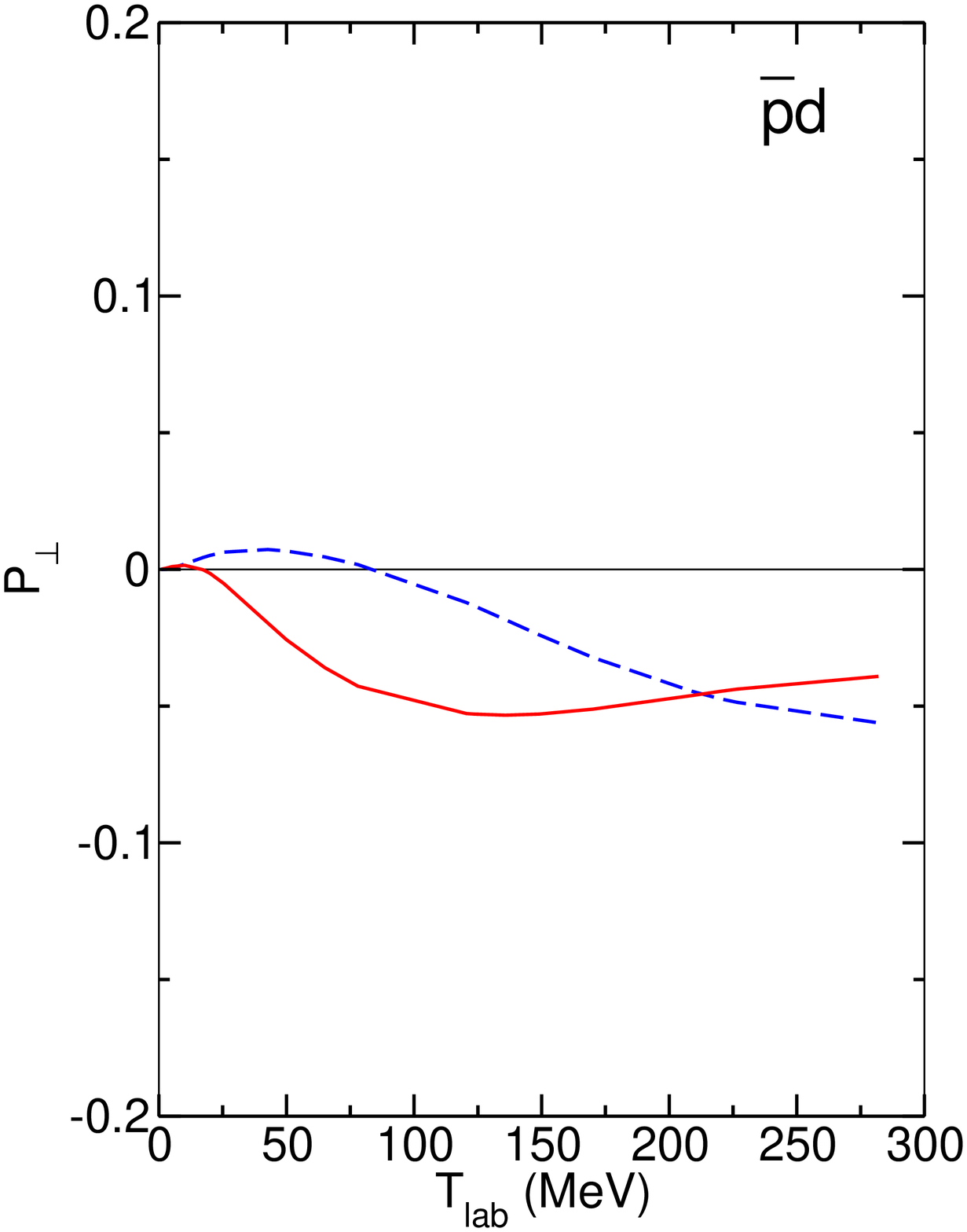}
\includegraphics[width=0.30\textwidth,angle=0]{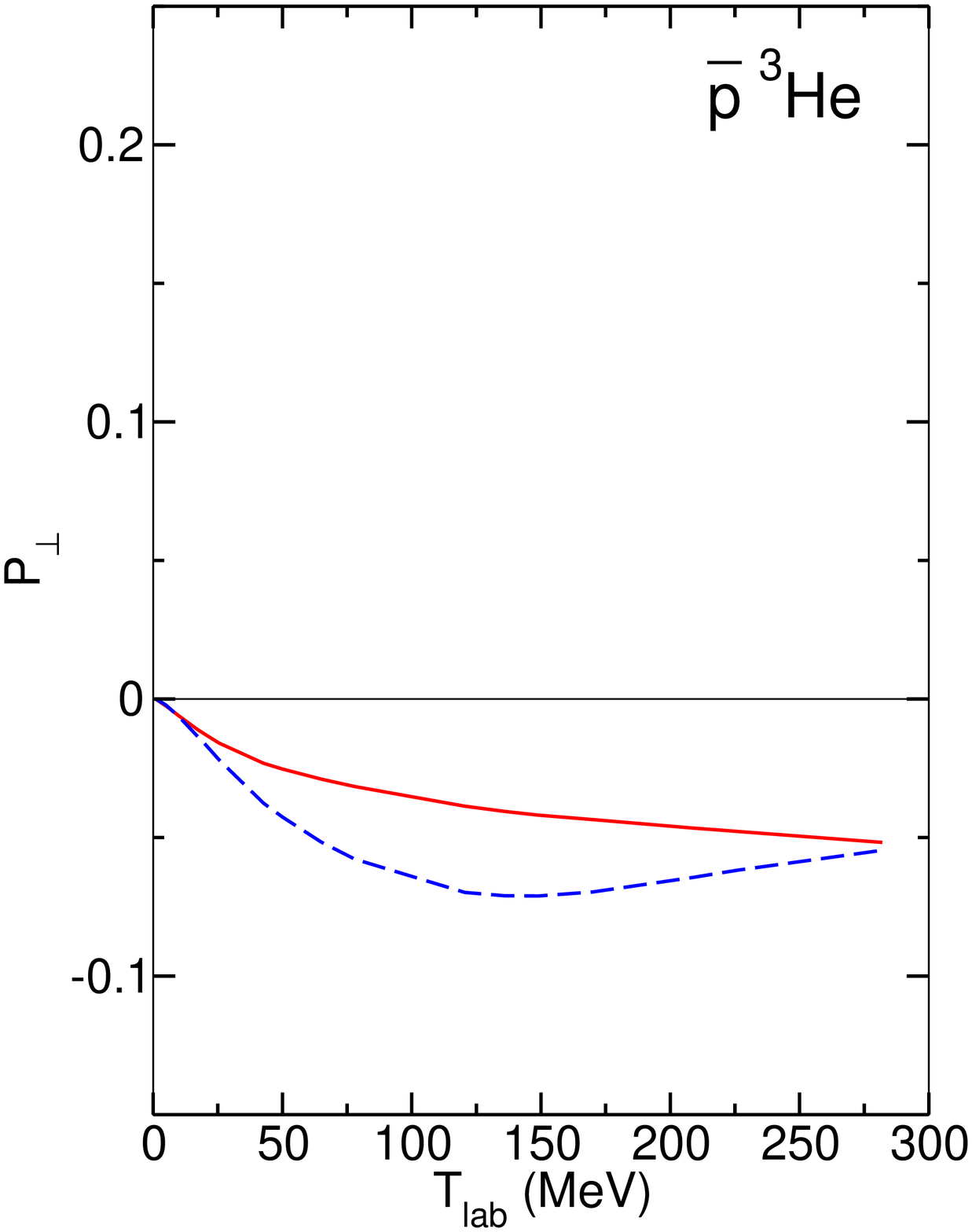}
\caption{
Dependence of the
transversal polarization $P_{\perp}$ (i.e. $P_{\bar p}(t_0)$ for
${\bfg \zeta}\cdot {\bf \hat k}=0$)
on the beam energy
for the target polarization  $P_T=1$ in the different reactions
$\bar pp$, $\bar pd$, and $\aphet$.
The results are for the models A (dashed line) and D (solid line).
The employed acceptance angle is $\theta_{acc}=10$ mrad. 
}
\label{pperp}
\end{figure}

According to the analysis of the kinetics of polarization \cite{MS,NNNP1},
the polarization buildup is determined mainly by the ratio of the polarized
total cross sections ($\sigma_1$, $\sigma_2$) to the unpolarized one ($\sigma_0$)
\cite{MS}. Let as define the unit vector ${\bfg \zeta}= {\bf P}_T/ P_T$,
 where $ {\bf P}_T={\bf P}_\tau$ is the  target polarization vector, which
  enters 
 Eq. (\ref{sigmatot}). The non-zero antiproton beam polarization vector
${\bf P}_{\bar p}$, produced  by the polarization buildup,
 is collinear to the vector ${\bfg \zeta}$ for any directions of ${\bf P}_T$
 and can be calculated from consideration of the kinetics of polarization.
 The general solution for the kinetic
 equation for $\bar p p$ scattering is given in Ref.~\cite{MS}. Here we
 assume that this solution is valid for $\aphet$ scattering too.
 Therefore, for the spin-filtering mechanism of the polarization buildup
 the polarization degree at the time $t$ is given
 by \cite{MS,DMS2}
\begin{equation}
P_{\bar p}(t)=\tanh\left [\frac {t}{2}(\Omega_{-}^{out}-
\Omega_{+}^{out})\right ],
\label{pdeg}
\end{equation}
where
\begin{equation}
\Omega_{\pm}^{out}=nf\left \{\sigma_0\pm P_T\left [\sigma_1 +
({\bfg \zeta}\cdot {\bf \hat k})^2\sigma_2\right ]\right \}.
\label{omega}
\end{equation}
Here $n$ is the areal density of the target and $f$ is the beam revolving
frequency. 
Assuming the condition $|\Omega_{-}^{out}-\Omega_{+}^{out}|
<< (\Omega_{-}^{out}+\Omega_{+}^{out}$), which was found in
Refs.~\cite{MS,DMS2} for $\bar p p$ scattering in storage rings at $n=10^{14}$
cm$^{-2}$ and $f=10^6$ c$^{-1}$,
one can simplify Eq. (\ref{pdeg}).
If one denotes the number of antiprotons in the beam at the time moment $t$
as $N(t)$, then the figure of merit is $P_{\bar p}^2(t)N(t)$. This value
is maximal at the time $t_0=2\tau$, where $\tau$ is the beam lifetime,
which is determined by the total cross section $\sigma_0$ of
the interaction of the antiprotons with the nuclear target,
\begin{equation}
\tau=\frac{1}{nf\sigma_0}.
\label{tau}
\end{equation}
To estimate the efficiency of the polarization buildup mechanism it is
instructive to calculate the polarization degree $P_{\bar p}$  at the time $t_0$
\cite{DMS2}. With our definition of $\sigma_1$ and $\sigma_2$ this quanitity is
given by 
\begin{eqnarray}
P_{\bar p} (t_0)&=&-2P_T\frac{\sigma_1}{\sigma_0}, \,\,\, if \,\,\,
{\bfg \zeta}\cdot {\bf \hat k}=0,
\nonumber \\
P_{\bar p} (t_0)&=&-2P_T\frac{\sigma_1+\sigma_2}{\sigma_0}, \,\,if\,\,\,
 |{\bfg \zeta}\cdot {\bf \hat k}|=1.
\label{RL}
\end{eqnarray}
 
Let us first look at the spin-dependent cross sections themselves which are 
presented in Fig.~\ref{Totals}. Note that here the corresponding calculations
are all done in the single-scattering approximation only, as
described in Sect. IIB and C. The c.m. acceptance angle used in
those calculations is $\theta_{acc}=10$ mrad. 
In principle, the corrections from multiple scattering to 
the spin-dependent cross sections could be worked out
by extending the formalism described in Refs.~\cite{Plat1}
to the $\aphet$ case. 
We expect that the multiple-scattering effects on those quantities
are roughly of the same magnitude (i.e. around 30 \% for energies above
20 MeV) as for the spin-independent cross sections. At least this was
found in case of $\bar p d$, reported in \cite{Sal11}.  
Therefore, we believe that the single-scattering approximation
provides a reasonable estimation for the magnitude of the
polarization-build-up effect in $\aphet$ scattering and we refrain 
from a thorough evaluation of the involved multiple-scattering effects in
the present analysis.
After all one has to keep in mind that the differences between the $\bar NN$
models A and D introduce significantly larger variations in the cross
sections $\sigma_1$ and $\sigma_2$, cf. Fig.~\ref{Totals}.

Our results suggest that the magnitude of the spin-dependent cross sections 
$\sigma_1$ and $\sigma_2$ for $\aphet$ are comparable to those for 
$\bar p p$ and $\bar p d$, at least as far as the hadronic part is concerned.
However, due to the larger charge of $^3\He$, Coulomb-nuclear interference 
effects turn out to be more important. Indeed, 
the Coulomb-nuclear interference cross sections $\sigma_i^{int}$ are
comparable to the corresponding polarized hadronic
cross sections $\sigma_1$  and $\sigma_2$ even at 100-200 MeV. 
 
The unpolarized cross section $\sigma_0^h$ (cf. left panel of 
Fig.~\ref{Totals}) is roughly a factor 3 larger than
the one for $\bar p p$ \cite{we09}, as expected. 
Moreover, the Coulomb cross section is significantly larger than in the
$\bar p p$ case. Indeed, the latter is still of similar magnitude 
as the purely hadronic cross section $\sigma_0^h$ at beam energies 
around 100 MeV. 

The polarization degree $P_{\bar p}(t_0)$ for ${\bfg \zeta}\cdot
{\bf \hat k}=1$ ($P_{||}$) at $P_T=P^d=1$ for $\aphet$ 
is shown in Fig.~\ref{ppara} versus the beam energy.
The results for ${\bfg \zeta}\cdot {\bf \hat k}=0$ ($P_\perp$) are displayed in 
Fig.~\ref{pperp}. For the ease of comparison the polarization degree for 
the $\bar p p$ and $\bar p d $ cases \cite{we11} are included too. 
The magnitudes of $P_{||}$ and $P_{\perp}$ in the region of the beam energy 0-300 MeV
are in the order of five percent. In case of $P_{||}$ they tend to be smaller than 
those predicted for $\bar p p$ \cite{DMS2,we11} and $\bar pd$ \cite{we11,Sal11} 
while for $P_{\perp}$ they are comparable to the ones for those other 
antiproton reactions. 

Since the polarization degree for $\bar p n$ was found to be in the order
of 20\% \cite{we11} one might naivly expect that it could be similar for 
$^3{\rm He}$ because, as mentioned above, in the latter the polarization is carried 
mainly by the neutron. However, the polarization degree is determined by the ratios 
of the 
spin-dependent cross sections $\sigma_i = \sigma_i^h + \sigma_i^{int}$ ($i=$ 1,2) to
$\sigma_0 = \sigma_0^h + \sigma_0^{int} + \sigma_0^C$, 
cf. Eq.~(\ref{RL}), and thus, is reduced by 
the larger unpolarized cross section $\sigma_0$ and, in particular, the larger
total Coulomb cross section $\sigma_0^C$ in the $\aphet$ system. 
In this context, note that also the beam lifetime decreases with 
increasing $\sigma_0$, see Eq.~(\ref{tau}). 

As discussed in Sect. III, if one goes beyond the single-scattering approximation 
the hadronic part of the unpolarized cross section 
$\sigma_0^h$ decreases by a factor of $\approx 1.4$ which, in principle, would 
lead to an increase of the polarization efficiency by the same factor. 
However, in case of $\bar p d$ it has been found that then also the spin-dependent
cross sections are reduced \cite{Sal11} by a similar amount so that there is 
practically no net effect. It is likely that the same will happen for 
$\aphet$ as well. 

\section{Summary}

In the present paper we employed two $\bar NN$ potential models 
developed by the J\"ulich group for a calculation of $\aphet$ and 
$\aphef$ scattering within the Glauber-Sitenko theory.
One of the aims was to examine in how far antiproton scattering 
off a polarized $^3\He$ target would be suitable for obtaining
a polarized antiproton beam via the spin-filtering method. 
The predicted spin-dependent cross sections for $\aphet$, 
evaluated in the single-scattering approximation for the J\"ulich $\bar NN$
models A and D, are comparable to those for the scattering of
antiprotons on polarized $^1{\rm H}$ or deuteron targets. 
However, since the total cross section is larger in case of $^3\He$ 
the resulting efficiency of the polarization buildup tend to be 
somewhat smaller than those for $\bar pp$ and $\bar p d$ so that 
one has to conclude that the use of a 
polarized $^3\He$ target might be less favorable for obtaining 
a polarized beam of antiprotons as required for the PAX experiment. 

Besides the issue of the polarization buildup for antiprotons,
$\aphet$ scattering is interesting for studying the spin dependence
of the elementary ${\bar p} N$ amplitudes. Since the spin-dependent 
part of $\aphet$ scattering is determined mainly by the $\bar p n$ 
amplitude, scattering of antiprotons on a polarized $\aphet$ target 
could reveal valuable additional information on this amplitude. 
It would supplement the constraints that could be provided by the
expected data on $\bar p d$ scattering from the AD experiment \cite{AD}, 
since in the latter a stronger interplay between the $\bar pp$ and $\bar p n$ 
amplitudes has to be expected.
Our results for unpolarized observables (integrated and differential 
cross sections) for $\aphet$ and $\aphef$, obtained within the 
Glauber-Sitenko approach, agree rather well with the available 
experimental information in the energy range from 20 MeV upwards. 
We view this as a strong indication that this formalism is 
suited for analyzing data for those reactions in the low- and
intermediate energy region. Of course, once concrete measurements with 
polarizated beam or target are planned, our calculations have to be improved 
and, specifically, corrections due to multiple scattering have to be also 
taken into account in the computation of polarization observables. 

\subsection*{Acknowledgements}
We acknowledge stimulating discussions with N.N.~Nikolaev
and F.~Rathmann. This work was
supported in part by the Heisenberg-Landau program.

\end{document}